\documentclass[a4paper,twocolumn,prb,showpacs,amssymb,aps]{revtex4}
\usepackage{dcolumn}
\usepackage{bm}
\usepackage{graphicx,color}
\newcommand{\tb}{$\mathrm{TbFe_3(BO_3)_4}$}
\newcommand{\re}{$\mathrm{RFe_3(BO_3)_4}$}
\newcommand{\gd}{$\mathrm{GdFe_3(BO_3)_4}$}
\newcommand{\nd}{$\mathrm{NdFe_3(BO_3)_4}$}
\newcommand{\mb}{$\mu_{\rm B}$}
\newcommand{\tn}{$T_{\mbox{\scriptsize N}}$}
\newcommand{\te}{$T_{\mbox{\scriptsize S}}$}
\newcommand{\bc}{$B_{\mbox{\scriptsize C}}$}
\newcommand{\tsr}{$T_{\mbox{\scriptsize SR}}$}
\newcommand{\figref}[1]{Fig.~\protect\ref{#1}}

\begin{document}

\title{Magnetization and specific heat of $\mathbf{TbFe_3(BO_3)_4}$: 
Experiment and crystal-field calculations}
\author{E.A. Popova$^1$}\author{D.V. Volkov$^1$}\author{A.N. Vasiliev$^1$}
\affiliation{$^1$Low Temperature Physics Department, Physics Faculty, Moscow State University,
119992 Moscow, Russia}
\author{A.A. Demidov$^2$}\author{N.P. Kolmakova$^{2,}$}\email[Electronic address: ]{npk@tu-bryansk.ru}
\affiliation{$^2$Bryansk State Technical University, 241035 Bryansk, Russia}
\author{I.A. Gudim$^3$}\author{L.N. Bezmaternykh$^3$}
\affiliation{$^3$L.V. Kirensky Institute of Physics, Siberian Branch of RAS, 660036 Krasnoyarsk,
Russia}
\author{N. Tristan$^4$}\email{n.tristan@ifw-dresden.de}\author{Yu. Skourski$^4$}\author{B. B{\"u}chner$^4$}\author{C. Hess$^4$}\author{R.
Klingeler$^4$} \affiliation{$^4$Leibniz-Institute for Solid State and Materials Research IFW
Dresden, 01171 Dresden, Germany}

\date{April 19, 2006; revised manuscript-\today}

\begin{abstract}
We have studied the thermodynamic properties of single-crystalline \tb. Magnetization measurements
have been carried out as a function of magnetic field (up to 50\,T) and temperature up to 350\,K
with the magnetic field both parallel  and perpendicular to the trigonal $c$-axis of the crystal.
The specific heat has been measured in the temperature range 2-300\,K with a magnetic field up to
9~T applied parallel to the $c$-axis. The data indicate a structural phase transition at 192~K and
antiferromagnetic spin ordering at $T_{\rm N}\approx 40$\,K. A Schottky anomaly is present in the
specific heat data around 20\,K, arising due to two low-lying energy levels of the Tb$^{3+}$ ions
being split by f-d coupling. Below \tn\ magnetic fields parallel to the $c$-axis drive a spin-flop
phase transition, which is associated with a large magnetization jump. The highly anisotropic
character of the magnetic susceptibility is ascribed mainly to the Ising-like behavior of the
Tb$^{3+}$ ions in the trigonal crystal field. We describe our results in the framework of an
unified approach which is based on mean-field approximation and crystal-field calculations.
\end{abstract}

 \maketitle

\section{Introduction}

The family of ferroborates with the general formula \re\ (R is a rare-earth ion or Y) attracts
considerable attention since its members exhibit a wide variety of phase transitions. They possess  magnetic structures which change as a function of temperature, external magnetic field and
substitutions in the rare-earth subsystem.\cite{Campa97,Hinatsu03,Balaev03,Levitin04} It was
recently discovered that \re , (R = Gd, Nd) exhibits multiferroic features, i.e. the coexistence
of elastic, magnetic and electric order parameters.\cite{Zvezdin05,Zvezdin06,Yen05} The various
ordering phenomena and their interaction cause anomalies in the dielectric permeability, electric
polarization and magnetostriction, both spontaneous and field-induced. Our study aims to
elucidate this interplay for the ferroborate \tb , in which both the rare-earth ions and the
Fe-ions form a magnetic subsystem.

\begin{figure}
\center{\includegraphics [width=1.0\columnwidth,clip] {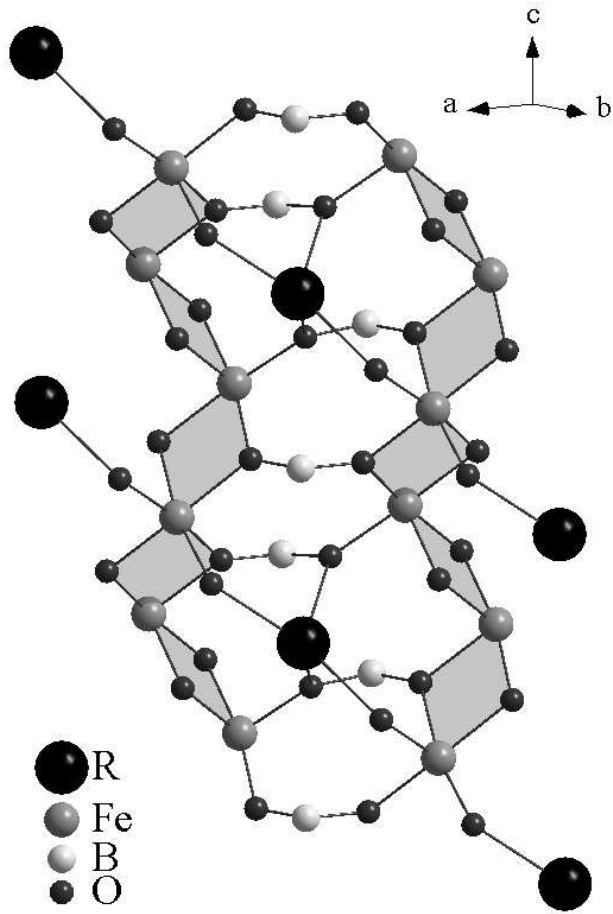}} \caption[] {\label{fig1} The crystal structure of \re. The spiral chains formed by FeO$_6$ octahedra along c-axis
(the shortest Fe-O-Fe exchange) are represented by the chains of shadowed areas.}
\end{figure}

The crystal structure of \tb\ at high temperatures is trigonal and belongs to the space group
$D_3^7$ (R32).\cite{Campa97, HTstructure,Klimin05} The main elements of the structure are spiral
chains of edge-sharing FeO$_6$ octahedra running along the $c$-axis. The rare-earth ions are
coordinated by triangular RO$_6$ prisms which are isolated from each other since they are
separated by regular BO$_3$ triangles and have no common oxygen ions. Both the BO$_3$ triangles
and RO$_6$ prisms connect three FeO$_6$ chains. There are no direct Fe-O-Fe links within the same
$c$-plane and the shortest interchain exchange paths are therefore given by Fe-O-R-O-Fe and
Fe-O-B-O-Fe, while the Fe-O-Fe exchange paths within the chains are much shorter (see
\figref{fig1}). In the case of \gd , the room temperature structure was found to be transformed to
the $D_3^4$ (P3$_1$2$_1$) space group at $T_S=156$\,K.\cite{Levitin04, Klimin05} Below $T_S$, two
non-equivalent Fe sites are present, which give rise to both a stretched and compressed
modification of the FeO$_6$ spiral chains.

\begin{figure}
\center{\includegraphics [width=1.0\columnwidth,clip] {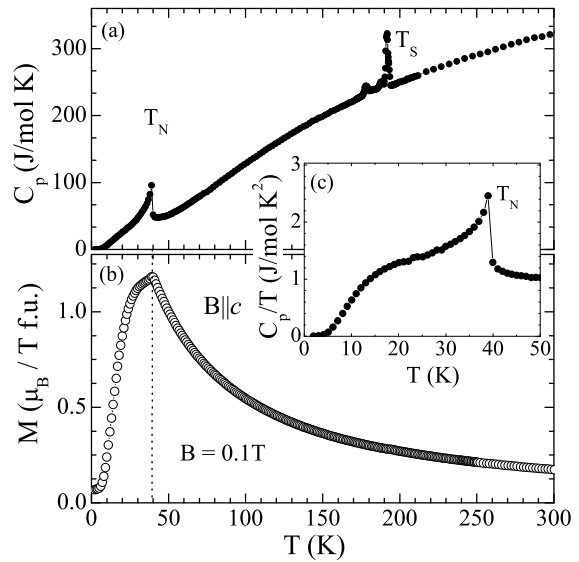}} \caption[]
{\label{fig2}Specific heat in zero magnetic field (a) and magnetization at $B=0.1$\,T (b) of \tb .
The inset (c) highlights the Schottky anomaly of $C_p/T$ at $T\sim 20$\,K. \te\ and \tn\ label the
(presumably) structural phase transition and the onset of antiferromagnetic spin ordering in the
Fe sublattice, respectively.}
\end{figure}

In \gd , the presence of different magnetic sublattices causes several magnetic transitions below
$T_S$, which are associated with antiferromagnetic spin ordering of the Fe-sublattice (\tn\
$\approx 37$\,K) and a spin reorientation at \tsr $\approx 9$\,K.\cite{Levitin04, Pankrats04,
Vasiliev06a, Vasiliev07} In addition, strong magnetostrictive and magnetoelectric effects below
\tsr\ are associated with a spin reorientation which occurs in external magnetic fields with
$B\|c$-axis.\cite{Zvezdin05,Zvezdin06,Yen05}

In this paper, we report on the thermodynamic properties of \tb . Our data on \tb\ show a
spin-flop phase transition, which is driven by magnetic fields $B||c$. This transition is
associated with a large jump of the magnetization. The highly anisotropic magnetic susceptibility
is ascribed mainly to the Ising-like behavior of the Tb$^{3+}$ ions in the trigonal crystal field.
Our experimental data are analyzed in the context of the unified approach which is based on
mean-field approximation and crystal-field calculations.
The spin-flop transition  is apparently accompanied by magnetoelastic effects. Magnetoelectric effects might therefore be associated with the spin-flop transition, leading \tb \ to be a candidate for a multiferroic compound.

\section{Experiment}

The single crystals of \tb\ were grown using a Bi$_2$Mo$_3$O$_{12}$-based flux.\cite{Bez04} The
seeds were obtained by spontaneous nucleation from the same flux. Single crystals were green in
color and had a good optical quality. The orientation of the crystals was performed by x-ray
diffraction. All magnetic measurements were performed with the external magnetic field either
parallel or perpendicular to the $c$-axis of the crystal. The AC-susceptibility at 1000 Hz and
DC-susceptibility were measured in magnetic fields of 0.001\,T and 0.1\,T, respectively, in the
temperature range 1.8-350\,K using a Quantum Design Physical Property Measurement System (PPMS).
The magnetization $M(B)$ was measured between 4.2 and 120\,K in fields up to 15\,T with a
vibrating sample magnetometer (VSM).\cite{Klingeler05} The field sweep rate was approximately
0.2\,T/min. In addition to this, a pulsed field magnetometer was used for magnetization
measurements up to 50\,T.\cite{Krug01} Here, the total pulse duration amounted to 0.05\,s. The
temperature dependence of the specific heat was measured in the temperature range 1.8-300\,K and
in magnetic fields up to 9\,T with a PPMS calorimeter.

\section{Experimental results\label{experiment}}

\tb\ exhibits two phase transitions below room temperature, as is illustrated by the specific
heat and magnetization data in \figref{fig2}. By comparison with \gd , the first order phase
transition at \te = 192\,K is tentatively attributed to a structural symmetry reduction to the low
symmetric P3$_1$2$_1$-phase.\cite{Klimin05} Actually, the data exhibit several first order
anomalies of the specific heat in the vicinity of  \te\ which are
probably related to structural changes. The origin of these anomalies can not be clarified by our
present study which focuses on the magnetic properties of \tb . The structural phase transition is
associated with an entropy jump of $\Delta S  \approx $ 1.9\,J/mol K. At \tn\ = 40 K, the onset of
antiferromagnetic spin order in the iron subsystem is demonstrated by a sharp jump in the specific
heat and a kink in the magnetization. We observe a specific heat jump of $\Delta C_p \approx $
59.4\,J/mol K. This result agrees with the jump $\Delta C_p$ which is predicted from the mean
field theory for the antiferromagnetic spin ordering of the $S=5/2$ Fe-sublattice:\cite{Morrish}
\begin{equation}
\Delta C_p = \frac{5S(S+1)}{S^2+(S+1)^2}R = 59\,\mbox{J/mol K} ,
\end{equation}
with $R$ being the gas constant. The entire spin entropy of the Fe$^{3+}$ subsystem appears to
develop only below \tn. The Fe$^{3+}$ subsystem must therefore be considered as a classical 3D
antiferromagnet, since short range spin correlations at significantly higher temperatures which
would be present if the Fe$^{3+}$ chains were able to form quasi one-dimensional magnets are
clearly absent. An analysis of the specific heat data \cite{Vasiliev06a} indeed confirms
that the magnetic ordering only occurs in the Fe subsystem while the Tb subsystem is
polarized by the Fe subsystem.

In contrast to the anomalies due to the structural and magnetic phase transitions at \te\ and \tn , respectively, the
Schottky anomaly in the specific heat at $\sim$20\,K reveals the temperature-driven population of
the ground state of the Tb$^{3+}$ ions split in the magnetic field of the Fe subsystem. No
spin reorientation (similar to the one seen in \gd\ ferroborate) was observed at low temperatures
in zero magnetic field.

\begin{figure}[h]
\center{\includegraphics [width=1.0\columnwidth,clip] {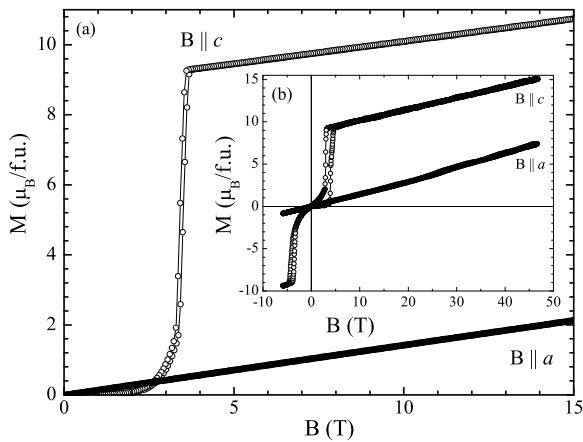}} \caption[]
{\label{fig3}Magnetization, at $T=4.2$\,K, for the magnetic field applied to the $c$- and
$a$-axis, respectively. Data obtained in a quasi-static field (VSM) are shown in (a) whereas (b)
displays pulsed magnetic field data (see text).}
\end{figure}

Depending on their direction, external magnetic fields can have a drastic effect on the magnetic
properties of \tb . As shown in \figref{fig3}, for applied magnetic fields parallel to the
$a$-axis there is a linear field dependence of the magnetization at 4.2\,K. The data imply
$\chi_a$=0.14\,$\mu_B$/T f.u. In contrast, applying a magnetic field along the easy magnetic
axis, i.e. the trigonal $c$-axis, causes a sharp jump in the magnetization at $B_c$(4.2\,K) $\sim$
3.5\,T. Here, the magnetic field drives a first order transition. We demonstrate further below
that the main feature of this transition is a spin-flop of the AFM ordered Fe$^{3+}$ spins which
is accompanied by the orientation of magnetic moments of the Tb$^{3+}$ subsystem along the external magnetic field. The
strongly discontinuous character of the transition is illustrated also by the pulsed field data in \figref{fig3}(b).
In the pulsed field ($\partial B/\partial t\sim 5000$\,T/s) the hysteresis at the
spin-flop transition is one order of magnitude larger than in the quasi-static measurement
($\partial B/\partial t\sim 10^{-3}$\,T/s) shown in \figref{fig3}(a), due to the relaxation processes and magnetocaloric effect. At high magnetic fields
$B>B_c$, the data show a linear $M(B)$ curve with $\chi_c$=0.13\,$\mu_B$/T f.u. being
slightly smaller than $\chi_a$.

The magnetization data in \figref{fig3} show a strong anisotropy. It is reasonable (and will be
shown in Sec. \ref{Comparison} in a detailed analysis) to attribute this anisotropy to the highly
anisotropic Tb$^{3+}$ magnetic moments, which are subject to the staggered field of the
antiferromagnetically ordered Fe spins. The discussion hence must not only address the two
subsystems of Fe and Tb moments, of which the Fe system divides in the two sublattices below \tn ,
but also the Tb subsystem must be discussed in terms of two sublattices since it is polarized by
the indirect exchange interaction with the iron subsystem.

The field dependences of the magnetization at 4.2\,K shown in \figref{fig3}, reveals two
anisotropy features: (1) A sharp increase of the magnetic moment $\sim$9\,\mb\ evolves at the
critical field \bc. (2) The linear contribution to $M(B||c)$ only exists above \bc . If the extra
moment (1) is neglected, the linear part of $M(B||c)$ is a straight line through the origin which
is very similar to $M(B||a)$. Such behaviour (2) is typical for an uniaxial antiferromagnet
magnetized along the easy axis. 

\begin{figure}
\center{\includegraphics [width=1.0\columnwidth] {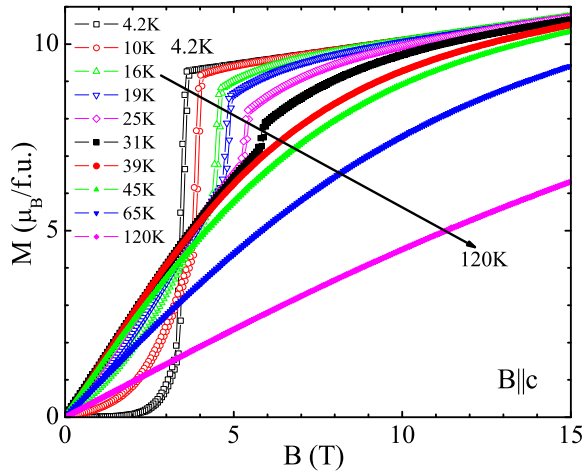}} \caption[] {\label{M_B_T} (Color online)  Magnetization for the magnetic field applied to the $c$-axis at various fixed
temperatures.}
\end{figure}

Upon heating (see Fig.~\ref{M_B_T}), the spin-flop transition shifts to higher temperatures and
vanishes above \tn , but it remains clearly visible in the entire antiferromagnetically spin
ordered phase. However, the magnetization jump reduces in value and becomes less sharp as it is
shifted towards higher fields. For fields larger than \bc\ as well as for $T > $\tn\ the
magnetization curves become nonlinear.

Below \tn\ and for fields $B\| c < $\bc , the staggered field of the Fe spins polarizes the
terbium subsystem due to $f$-$d$ interactions yielding two Tb sublattices with the magnetic
moments oriented oppositely and along the $c$-axis. For $B||c$ the effective field acting on the
terbium sublattice which magnetic moment opposes the applied field is decreased as the applied
field increases and this magnetic moment tends to diminish. This is a factor determining the
character of the magnetization at 4.2\,K prior to the phase transition, since at low temperatures
the longitudinal susceptibility of the iron subsystem  is very small. At \bc\, the staggered field
acting on the Tb$^{3+}$ magnetic moments vanishes as the magnetic moments of Fe$^{3+}$ ions become
oriented nearly perpendicular to them. As a result, the magnetic moments of Tb$^{3+}$ ions align
with the external magnetic field reaching the saturation value of about 9 $\mu_B$ at low
temperatures.

In the spin-flop phase both terbium sublattices have magnetic moments directed along the field
$B||c$ and the magnetic moments of iron sublattices bend towards the field direction. This part of
the magnetization curve allows us to estimate the transverse susceptibility of the iron subsystem:
$\chi_\perp ^{Fe}\approx \chi_c$ = 0.13\, $\mu_B$/T f.u. The observation of $\chi_a>\chi_c$
suggests a small contribution from the Tb subsystem for $B||a$. At higher temperatures, as shown in Fig. 4, the
magnetization curves gradually become less sharp. In the initial collinear phase the iron
subsystem begins to contribute to the magnetization as longitudinal susceptibility grows with
increasing temperature, thus stabilizing the initial phase.

\begin{figure}
\center{\includegraphics [width=1.0\columnwidth] {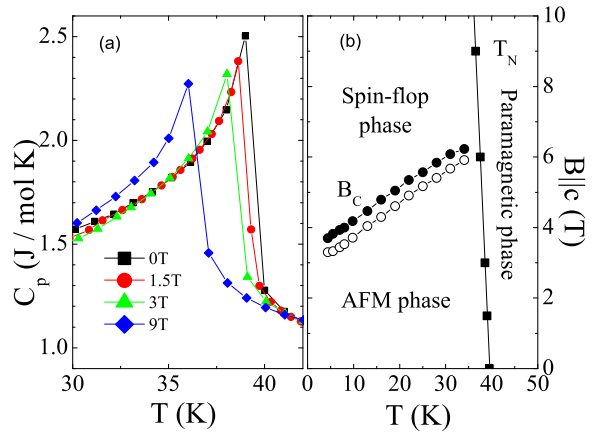}} \caption[] {\label{phase} (Color online) (a) Specific heat anomaly at \tn\ for different magnetic fields and (b) Magnetic phase
diagram for $B\|c$ from magnetization and specific heat data. \bc\ labels the critical field of
the spin-flop transition. Full (open) data markers refer to increasing (decreasing) quasi-static
fields, i.e. $\partial B/\partial t\sim 10^{-3}$\,T/s (see the text).}
\end{figure}

The magnetic phase diagram, constructed from this data and from specific heat measurements in
magnetic fields, is presented in \figref{phase}. The difference in phase boundaries shown by open
and full symbols accounts for the hysteretic phenomena and finite slope of M(B) curves at the
spin-flop transition. This diagram shows that magnetic fields parallel to the $c$-axis slightly
suppress the antiferromagnetic spin ordering temperature of the Fe sublattice. In contrast, the
critical field of the spin-flop transition increases upon heating.

\section{Theory\label{theory}}

The magnetic properties of terbium ferroborate are governed by both magnetic subsystems
interacting with each other. As already mentioned above, the crystal structure of rare-earth
ferroborates\cite{Campa97,Hinatsu03,Klimin05,HTstructure} suggests that the main feature of their
magnetic structure is formed by the helical iron chains along the $c$-axis. However, our
experimental data provide no evidence for quasi-1D magnetic properties, which indicates
significant interactions between the chains. Moreover, the experimental magnetization curves and 
phase diagram have the form which is typical for 3D uniaxial antiferromagnets. This fact provides 
reason enough to consider the magnetic structure of the Fe subsystem as that of iniaxial aitiferromagnet 
with two sublattices magnetic moments of which are equal in the absence of field. The interactions hierarchy in the 
system is such that it stabilizes the orientations of Fe and Tb magnetic moments along the trigonal axis. 
Indeed, the strongest interaction in the compound is the crystal field (the splitting of the ground 
multiplet comprises about 400-500 \,cm$^{-1}$, the energy interval between the ground quasi-doublet and 
the nearest excited states is of the order of 200 \,cm$^{-1}$), and the non-Kramers Tb$^{3+}$ ion in the 
trigonal crystal field acquires a huge anisotropy ($g_c$$\sim$  18\,, $g_a$ $\sim$ 0\,) and becomes an Ising ion with 
the preferred orientation of magnetic moment along the trigonal axis. Interaction comparable in value 
with the crystal field is absent in the system, therefore there is no source of any other orientation of 
Tb magnetic moments. And so there are grounds to suggest a collinear arrangement of magnetic moments 
in \tb\ . We note however that only neutron diffraction studies on an 11B-enriched compound can 
unambiguously resolve the magnetic structure under discussion.       

In the following section we model the magnetic properties of \tb\ using molecular field theory.
Within this approximation the total Hamiltonian, which comprises
the Hamiltonians of the iron and the terbium subsystems
as well as the one considering the Tb-Fe interaction,
 can be expressed as a sum of single-particle
Hamiltonians.  In the presence of an external magnetic field $B$, the
effective Hamiltonians of the Tb/Fe ion of the $i$-th sublattice ($i$=1 and 2) can be written as:
\begin{equation}
{\label{HamTb}} {\cal H}{_i}(\textrm{Tb}) = {\cal H}{^i}_{CF} - g_{J}{\mu}_{B} \textbf{J}_{i}
[\textbf{B} + \textbf{B}_{mi} (\textrm{Tb})],
\end{equation}
\begin{equation}
{\label{HamFe}} {\cal H}{_i}(\textrm{Fe}) = - g_{S}{\mu}_{B} \textbf{S}_{i} [\textbf{B} +
\textbf{B}_{mi} (\textrm{Fe})],
\end{equation}
where $g_{J}$ is the Lande factor and $\textbf{J}_{i}$ is the angular momentum operator of the
rare-earth ion; $g_{S}=2$ is the g-factor and $\textbf{S}_{i}$ is the spin momentum operator of
the Fe ion; $\textbf{B}_{mi}$(Tb/Fe) are the molecular fields acting on the Tb or Fe ion in the
$i$-th sublattice. The crystal-field Hamiltonian ${\cal H}_{CF}$ is governed by the symmetry of the
local environment of the rare-earth ion.

As explained in Sec.~\ref{experiment}, from the analogy to the structural phase transition at
156~K in \gd\ we suppose that the specific heat anomaly at \te\ = 192\,K reflects a crystal
symmetry reduction from the trigonal space group R32 to the  trigonal group P3$_1$2$_1$.
This implies that the local symmetry of the Tb$^{3+}$ ion is reduced from $D_3$ at $T>T_S$ to
$C_2$ at $T<$ \te . Nevertheless (as will be discussed below) a simpler Hamiltonian of $D_3$
symmetry is sufficient for describing the low-temperature magnetic properties of \tb, since the
available crystal-field parameters of isostructural compounds give rise to two closely-spaced
(with a splitting less than 2-3 cm$^{-1}$) singlets as a low-lying state of the ground multiplet.
These singlets are responsible for the Ising-like behavior of this non-Kramers ion in both the
paramagnetic and the ordered phases.

In order to describe the thermodynamic properties of rare-earth compounds we only need to consider
the ground multiplet. In terms of the equivalent operators $O^m_n $ the crystal field Hamiltonian ${\cal H}_{CF}$ reads as following:
\begin{eqnarray}
\label{HamCF} {\cal H}{_{CF}} &=& \alpha_J B^0_2 O^0_2 + \beta_J( B^0_4 O^0_4 + B^3_4 O^3_4)\\
&& +\gamma_J (B^0_6 O^0_6 + B^3_6 O^3_6 + B^6_6 O^6_6) ,\nonumber
\end{eqnarray}
where $B^m_n$ are the crystal-field parameters for the $D_3$ symmetry, $\alpha_J$, $\beta_J$ and
$\gamma_J$ are the Stevens coefficients. The former are unknown for the Tb$^{3+}$ ion in \tb .
However, our results depend only weakly on the actual crystal-field parameters since the magnetic
properties of \tb\ are essentially controlled by the magnetic behavior of the iron subsystem and
by the Ising character of the non-Kramers Tb$^{3+}$ ion in the crystal field of the trigonal
symmetry. We hence could not determine the crystal-field parameters from the experimental data.
Instead, we have used the crystal-field parameters of the isostructural compounds, in particular
the rare-earth aluminoborates (see e.g. Ref.~\onlinecite{Cascales01}) and refined these using the
data for crystal-field splitting of Nd$^{3+}$ in \nd.~\cite{Chukalina03}

We have written the expressions for the molecular fields acting on the Tb and Fe ions from the
$i$-th sublattice on the basis of the interactions hierarchy and magnetic structure of \tb . A
detailed analysis of the interactions which was performed for the low temperature phase of \gd\ in
Ref.~\onlinecite{Pankrats04} also seems to be quite applicable to \tb . According to this
analysis, a rare-earth ion does not interact with iron ions from the same c-plane but interacts
antiferromagnetically with iron ions from the neighboring planes, it is clearly shown in Fig.7b of Ref.12. The interaction
between rare-earth ions can be neglected because the distances between them are larger than 6~\AA\
and no reasonably superexchange path could be introduced. Consequently, none of the ferroborates
demonstrates an intrinsic ordering in the rare-earth subsystem. The molecular fields
$\textbf{B}_{mi}$(Tb) and $\textbf{B}_{mi}$(Fe) can therefore be written as
\begin{equation}
{\label{HmiTb}} {\textbf{B}_{mi}}(\textrm{Tb}) = \lambda_{fd}
\textbf{M}_{i},
\end{equation}
\begin{equation}
{\label{HmiFe}}{\textbf{B}_{mi}}(\textrm{Fe}) = \lambda \textbf{M}_{j} + \lambda_{fd}
\textbf{m}_{i},\ j=1\ \textrm{or}\ 2,\ j\neq i,
\end{equation}
where $\lambda_{fd}<0$ and $\lambda<0$ are the molecular constants of the Tb-Fe and Fe-Fe
antiferromagnetic interactions. The magnetic moments $\textbf{M}_i$ and $\textbf{m}_i$ of the
$i$-th iron and terbium  sublattices are defined as
\begin{equation}
{\textbf{M}_i= 3g_S \mu_B\langle{\textbf{S}_i}\rangle}\ \textrm{and}\ {\textbf{m}_i = g_J \mu_B
\langle{\textbf{J}_i}\rangle}.
\end{equation}

The thermodynamic potential of the system (per formula unit) has the following form:
\begin{eqnarray}
{\label{PhiTH}}\Phi(T,B)&=&\frac{1}{2}[-k_BT\sum_{i=1}^2\ln Z_i(\textrm{Tb})\\
&& + \frac{1}{2}\sum_{i=1}^2g_J\mu_B \langle{\textbf{J}_i}\rangle \textbf{B}_{mi}(\textrm{Tb})\nonumber\\
&& - 3k_BT\sum_{i=1}^2\ln Z_i(\textrm{Fe})\nonumber\\
&&+ \frac{1}{2}\sum_{i=1}^2 3g_S\mu_B
\langle{\textbf{S}_i}\rangle \textbf{B}_{mi}(\textrm{Fe)}], \nonumber
\end{eqnarray}
with the partition functions 
\begin{equation}
{Z_i(Tb/Fe)=\sum_{n}   \exp[\frac{-{E}_{ni}(Tb/Fe)}{k_BT}]},
\end{equation}
where ${E}_{ni}(Tb/Fe)$ are the eigenvalues of corresponding Hamiltonians (Eqs.2 and 3).
To find the magnetic moments of the Tb and Fe subsystems Eqs.7 in the mean-field
approximation one has to solve the self-consistent problem of deducing their values and
orientations on the basis of Hamiltonians Eqs.~\ref{HamTb} and ~\ref{HamFe} while considering the
condition of the minimum of the thermodynamic potential (Eq.~\ref{PhiTH}) for a given temperature
and field. The right part of equation for $M_i$ is the relevant Brillouin function, as it should be in the case of
an equidistant spectrum which is typical for the Fe$^{3+}$ ion with an orbital singlet as a ground
state (S-ion). Substituting the magnetic moments for particular phases into the thermodynamic
potential (Eq. \ref{PhiTH}), we obtain the energies of the phases and find the critical fields for
the phase transitions from the condition of their equality. The magnetization of the compound (per
formula unit) is given by:

\begin{equation}
{\label{Magnet}}{\textbf{M}=\frac{1}{2} \sum_{i=1}^2(\textbf{M}_i+\textbf{m}_i)}.
\end{equation}

Both magnetic subsystems contribute to the initial magnetic susceptibility of terbium ferroborate:
\begin{equation}
{\chi_i = \chi_i^{\textrm{Fe}} + \chi_i^{\textrm{Tb}}},\ i = \parallel\ \textrm{or}\ \perp.
\end{equation}
In the paramagnetic range, where the interaction between iron and terbium subsystems can be
neglected, the magnetic susceptibility of the Tb subsystem can be calculated with the help of the
well-known Van Vleck formula on the basis of the crystal-field Hamiltonian Eq. \ref{HamCF}. For the ordered phase at
$T<T_N$ as well as for the paramagnetic phase, the initial magnetic susceptibilities of the
compound can easily be found from the initial linear parts of the magnetization curve $M(B)$
calculated for the field along and perpendicular to the trigonal axis. 

The contribution of the Tb$^{3+}$ subsystem to the heat capacity of the \tb\ compound can be
calculated with the help of the usual expression (per rare-earth ion, i.e. per formula unit):
\begin{equation}
{C_{Tb} = k_B \frac {\langle E^2 \rangle - \langle E \rangle ^2}{(k_BT)^2}               }.
\end{equation}
The thermal averages were calculated for the spectrum of Tb$^{3+}$ ion formed by the crystal
field and by interactions with the iron subsystem and an external magnetic field.

\section{Comparison of experimental data and theoretical calculations\label{Comparison}}

\subsection{Magnetization}

\begin{figure}
\includegraphics [width=1.0\columnwidth,clip] {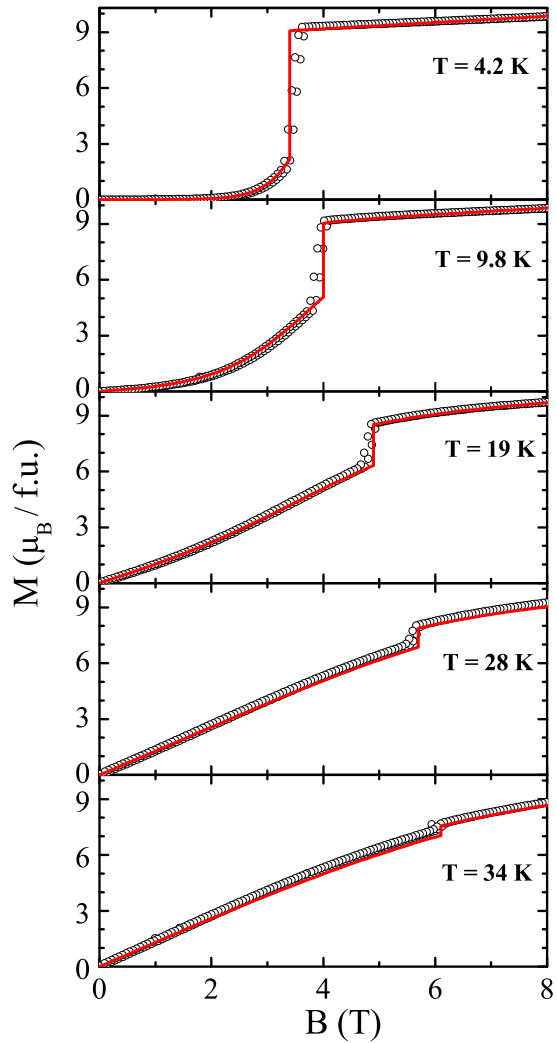}
\caption{\label{magtemp} (Color online) Experimental (symbols) and calculated (lines) magnetization
curves of \tb\ for various temperatures with $B\parallel c$.}
\end{figure}

In order to carry out a quantitative analysis of the magnetization curves for \tb\ in accordance
with the approach presented in the previous section, the magnetizations in the collinear and flop
phases were calculated in the mean-field approximation and the critical fields of the phase
transitions \bc \ were found from the equality of their thermodynamic potentials. In
\figref{magtemp} the experimental and calculated magnetization curves are displayed for several
temperatures.

The analysis of the magnetization curve at 4.2~K provides a possibility to determine some
parameters of the compound. The f-d interaction parameter $\lambda_{fd} = -0.253$\,T/$\mu_B$ was
calculated by fitting the initial part of $M(B)$ where the contribution of Fe subsystem is negligibly small because of its small longitudinal susceptibility at low temperatures. This value of $\lambda_{fd}$ is in good
accordance with the experimental data for the low-temperature splitting $\Delta$ of low-lying
singlets of the Tb$^{3+}$ ion in \tb\, which is around 32\,cm$^{-1}$.\cite{Privat06} Reasonable
crystal-field parameters for Tb$^{3+}$ in \tb\ yield the g-tensor component along the trigonal
axis in the range from 17.5 to 17.8, resulting in a value of $\Delta(T=0 \,\textrm{K}) = \mu_B g_z
B_{fd}(T=0  \,\textrm{K} ) = \mu_B g_z |\lambda_{fd}|M(T=0 \,\textrm{K} )\approx 31$\,cm$^{-1}$.

The value of the exchange Fe-Fe parameter, responsible for the transverse susceptibility,
$\lambda_{1}=-3$\,T/$\mu_B$, has been obtained from the slope of the $M(B)$ dependence measured at
$T=4.2$\,K in the flop phase. However, if the exchange parameter for the collinear phase is
taken equal to this value then the calculated phase transition critical field exceeds the
experimentally observed value and a similar relation between experimental and calculated critical
fields takes place for all temperatures. On the other hand, if one takes the exchange parameter in the collinear phase to be $\sim1\%$ less
than that in the flop phase, the energy of the collinear phase increases slightly, the critical
field decreases and its calculated value is in a good agreement with the experimental one (see
\figref{magtemp}). The difference between the values of the exchange parameter $\lambda$ for two
phases could be explained by the presence of the magnetoelastic effects accompanying the
field-induced phase transition. In support of this hypothesis, it is known\cite{Zvezdin05} that,
in \gd\ with the $S$-ion Gd$^{3+}$, the field-induced phase transition in the low-temperature
phase ($T<10$\,K) results in a magnetostriction jump of the order of $2\cdot10^{-5}$. In \tb\ the highly
anisotropic terbium ion should give a much more  essential contribution to the magnetostriction at the
field-induced spin-flop transition than the S-ion Gd$^{3+}$. Indeed, calculations of the multipole moments $Q_{nm}=\alpha_n
\langle O_n^m \rangle$ of the Tb$^{3+}$ ion in \tb\ (which provide an information on the
rare-earth contribution to the magnetostriction\cite{Kolmakova90}) have shown that their changes
at the spin-flop phase transition are very significant. 

\begin{figure}
\includegraphics [width=1\columnwidth,clip] {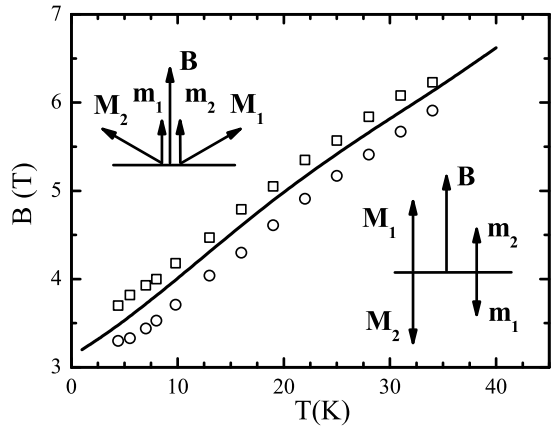}
\caption{\label{phasdiag}The $B-T$ phase diagram for the spin-flop transition in \tb\ at
$B\parallel c$ (cf. \figref{phase}b ). The symbols and the line represent the experimental data
and the calculated phase boundary, respectively; the difference in experimental phase boundaries reflects a finite slope of M(B) curve caused by a sample shape and hysteresis phenomena.}
\end{figure}

To analyze the temperature evolution of the magnetization we have to take into account the
exchange between the Fe chains which is responsible for the three-dimensional ordering. This value
of $\lambda$ enters the Brillouin function in Eq.7, the well-known explicit form of which is not given for the sake of brevity. It was chosen from the best agreement between experimental and calculated magnetization
curves for all temperatures as $\lambda_2=-2$\,T/$\mu_B$. The $B-T$ phase diagram for the
spin-flop transition (see \figref{phasdiag}) demonstrates a good agreement between the
calculated and experimental data. We emphasize that quantitative description of the magnetization isotherms of \tb\ in the
mean-field approximation requires an introduction of two exchange parameters characterizing Fe-Fe
antiferromagnetic interaction, $\lambda_1$ and  $\lambda_2$. The $\lambda_1$ enters Eq.
\ref{PhiTH} for the energy and determines the critical field of the spin-flop transition $B_{C}$
and the transverse susceptibility of the iron subsystem in the flop phase. The $\lambda_2$ determines
 the magnitude of iron magnetic moments $M_i$ for specific temperature and
field as well as the Neel temperature. The necessity of introducing two  exchange parameters is a
consequence of the chain structure of the ferroborate when considered in the mean-field
approximation. The \textit{intra}chain exchange interactions are described by $\lambda_1$, whereas
the \textit{inter}chain exchange interactions (which are responsible for the three-dimensional
magnetic order) are described by $\lambda_2$. Our experimental data
and the results of our theoretical analysis indicate that these two quantities are of the same
order of magnitude.

\subsection{AC magnetic susceptibility}

\begin{figure}
\includegraphics [width=1\columnwidth,clip] {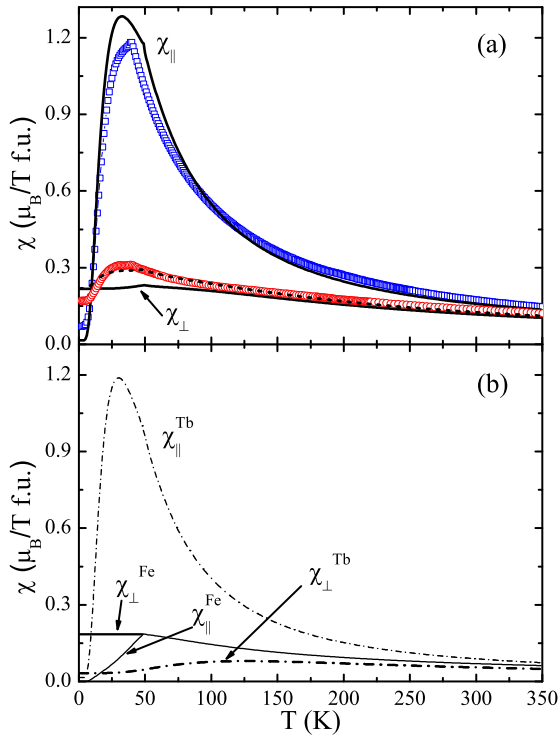}
\caption{\label{fitmag} (Color online) Temperature dependencies of the initial magnetic
susceptibility in \tb\ along ($\chi_{\parallel}$) and perpendicular ($\chi_{\perp}$) to the
trigonal axis. (a) Symbols are experimental points; solid thick lines are calculated curves for
$\chi_{\parallel}$ and $\chi_{\perp}$; thick dashed line represents calculation of
$\chi_{\perp}$ for the misorientation angle with regard to the basal plane in 3$^{\circ}$.
(b) Thin solid and dot-dashed lines are calculated curves for
iron and terbium subsystems, respectively.}
\end{figure}

\begin{figure}
\includegraphics [width=1\columnwidth,clip] {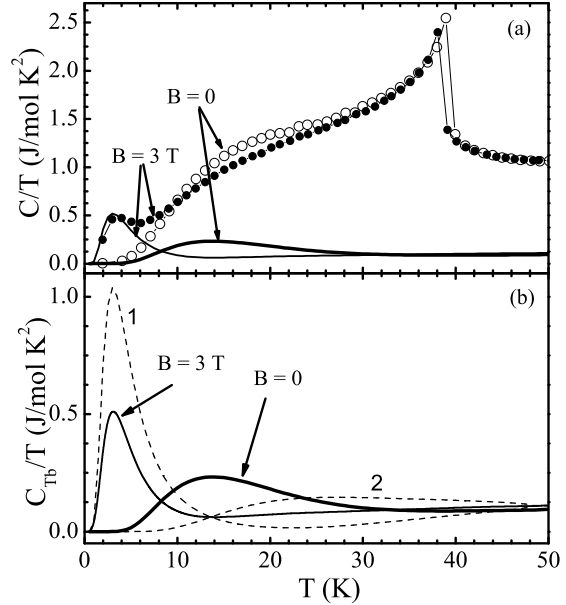}
\caption{\label{specheat}(a) Specific heat of \tb : open and closed symbols represent experimental
data at $B=0$ and 3\,T respectively. Thick and thin solid lines correspond to calculations of the Tb contributions for
$B=0$ and 3\,T respectively.  (b) Calculated specific heat of the Tb subsystem: dashed lines 1 and
2 show calculated contributions for different sublattices at $B=3$\,T.}
\end{figure}

As has already been discussed above, the strong anisotropy of the magnetic susceptibility is
caused by the contribution of the Tb$^{3+}$ ion with its Ising character of magnetic behavior. The
$\chi(T)$ dependencies have been calculated using parameters of \tb\ derived from the analysis of
the magnetization isotherms. However, as one can see in \figref{fitmag} the experimentally
obtained and calculated $\chi_{\perp}(T)$ curves differ substantially.
The non-zero susceptibility $\chi_\parallel$ at very low temperatures and a growth of $\chi_\perp$
upon increasing the temperature up to \tn\ could be ascribed to a misorientation of the sample by
a few degrees with respect to the crystallographic axis. In fact, the inclusion of a
misorientation by 3$^{\circ}$ with respect to the basal plane leads to a significantly better
agreement between the calculated and experimental data. The variation of the crystal-field
parameters within reasonable limits does not result in  any significant changes in the description
of the $\chi(T)$ dependencies; the agreement between experimental and calculated curves remains
roughly the same.

A distinctive convexity of the experimental $\chi_{\parallel}(T)$ curve in the temperature range
30-35\,K is associated with a Schottky-type anomaly, i.e. with the change of  occupation of two
low-lying terbium singlets. This anomaly can also be observed in the calculated curves, especially
in the susceptibility of the Tb subsystem. The calculated Neel temperature is higher than the
corresponding experimental value but this is an usual feature of the mean-field theory.

\subsection{Heat capacity}

In the following, the specific heat and the effect of the magnetic field will be analyzed in terms
of our approximation. The experimental specific heat data in magnetic fields $B=0$ and 3\,T along
the $c$-axis are presented in $C/T$ vs. $T$ coordinates in \figref{specheat}a. As one can see, the
magnetic field shifts the Schottky anomaly to lower temperatures, specifically  for $C(T)$ from
$\sim$ 20\,K at $B=0$ to $\sim$ 5\,K at $B=3$\,T. The anomaly at around 5\,K is particularly
evident since in this temperature range both the phonon and magnetic contributions are small.

\begin{figure}
\includegraphics [width=1\columnwidth,clip] {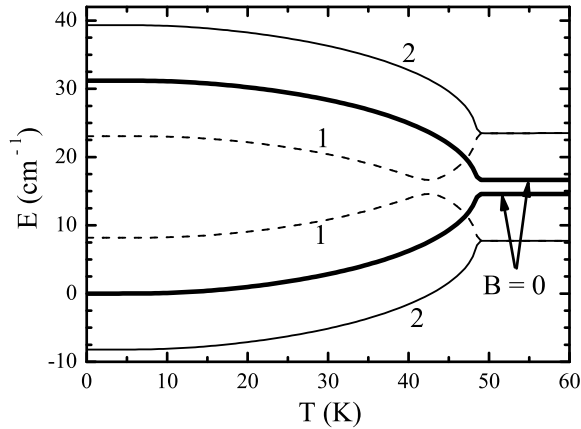}
\caption{\label{split}Temperature dependencies of the splitting of the ground quasi-doublet of the
Tb$^{3+}$ ion in \tb\ at $B=0$ and 3\,T  for sublattices with magnetic moments (1) opposite to the
field direction and (2) along the field.}
\end{figure}

The calculated temperature dependencies of the Tb contribution to the specific heat are presented
in \figref{specheat}b. The calculated $C_{\textrm{Tb}}(T)$ for $B=0$ demonstrates the Schottky
anomaly around 20\,K. As has already been mentioned above, the anomaly is caused by the occupation
of the singlet level separated from the ground singlet by the temperature-dependent splitting due
to the Tb-Fe interaction (\figref{split}).

Our analysis in \figref{specheat}b suggests that the Tb contribution to the specific heat is
different for the two Tb sublattices if an external magnetic field is applied along the $c$-axis.
For the terbium sublattice with magnetic moments opposed to the field direction, the Schottky
anomaly shifts to lower temperatures. This is easy to understand because an increase of the
magnetic field leads to a decrease of the energy gap between two singlets (see \figref{split}). It
means that an external field demagnetizes this terbium sublattice. For the other terbium
sublattice, however, the energy gap between the two singlets increases, the increased external
field magnetizes the sublattice and the Schottky anomaly moves to higher temperatures. We
emphasize the agreement between the experimental and numerical data which confirms that the
description of Tb-Fe coupling in \tb\ provided in this work is appropriate.

\section{Conclusions}

Our measurements of the specific heat and the magnetization and the theoretical description of the
experimental data provide a detailed understanding of the thermodynamical properties of \tb . Our
unified approach describes the presence of the spin-flop transition for magnetic fields applied
parallel to the trigonal $c$-axis and the related magnetization jump below \tn . Microscopically,
the highly anisotropic magnetization data have been attributed to the Ising character of the
Tb$^{3+}$ ion in the trigonal crystal field. In addition, the presence of a Schottky anomaly in
the specific heat data around 20\,K and its shift as a function of the applied magnetic field have
been explained as the Tb contribution to specific heat in terms of two low-lying energy levels of the Tb$^{3+}$ ions being split by f-d
coupling.

In particular, our approximation considers the exchange interaction of the Fe-subsystem as well as
the coupling between Fe- and Tb-subsystems. The $f-d$ coupling parameter has been determined and
its value is in good agreement with available spectroscopic data. The antiferromagnetic exchange
interaction in the iron subsystem has been described with the help of two parameters. The parameter $\lambda_1$ is responsible for bending the iron magnetic moments  in the flop phase, which occurs against the intrachain exchange, and determines the spin-flop transition field; the parameter $\lambda_2$ is connected with the interchain exchange interaction responsible for the three-dimensional ordering, it controls the Neel temperature and magnitudes of the iron magnetic moments for specific temperatures and fields. The relation $|\lambda_1| > |\lambda_2|$ is naturally fulfilled because the Fe-O-Fe exchange paths within the chains are much shorter than the interchain exchange paths. The appearance of two exchange parameters is a special feature of consideration of magnetic properties of the compound with chain structure in the framework of the mean-field approximation, which is known to be of limited validity for description of magnetic lattices of lowered dimensionality. A substantial influence of the
magnetoelastic effects which accompany the field-induced phase transition on the value of the
molecular parameter responsible for bending the iron magnetic moments in the flop phase has been
found.

\begin{acknowledgments}
This work was partially supported by the Deutsche Forschungsgemeinschaft through 436 RUS
113/864/0-1, FOR 520(A7) and RFBR Grants 06-02-16088, 07-02-00350. A.A.D. acknowledges support through Russian Federation's President Grant MK-4393.2006.2. 
\end{acknowledgments}

\end{document}